\documentclass[a4]{article}
\usepackage{color, graphicx}
\usepackage{amsmath}
\usepackage{amssymb}
\usepackage[colorlinks=true, urlcolor=blue]{hyperref}
\begin{document}

\noindent
\begin{center}
\begin{LARGE}
\textbf{{Evolution equations of the truncated moments of the parton densities.
A possible application}}\\
\end{LARGE}
\vspace{10mm}
\begin{Large}
{Dorota Kotlorz\footnote{Division of Physics, Ozimska 75, 45-370 Opole,
Poland, e-mail: {\tt d.strozik-kotlorz@po.opole.pl}}
and Andrzej Kotlorz\footnote{Division of Mathematics, Luboszycka 3,
45-036 Opole, Poland}
\\Opole University of Technology}
\end{Large}
\end{center}

\abstract{
A possible application of the evolution equation for the truncated Mellin
moments to determination of the parton distributions in the nucleon is
presented. We find that the reconstruction of the initial parton densities
at scale $Q_0^2$ from their truncated moments at a given scale $Q^2$ is exact
and unique for small number of free parameters ($\leq 3$), even for the
limited $x$-region of experimental data. For larger number of adjustable
parameters the obtained fits are not unique and one needs an additional
knowledge of the small-$x$ behaviour of the parton densities to make the
reconstruction procedure reliable. We apply successfully our method to
HERMES and COMPASS spin-dependent valence quark data.
}
\\ \\
PACS {12.38.Bx} {Perturbative calculations}, {11.55.Hx},
{Truncated moments}

\section{Introduction}
\label{intro}

Determination of unpolarised as well polarised parton distribution
functions (PDFs) is nowadays a topic of intensive theoretical and
experimental investigations.
At the dawn of the LHC, where high center-of-mass energy
$\sqrt (s)=14\;{\rm TeV}$ is available and all large $Q^2$ reactions are
parton collisions, the precise knowledge of the PDFs is needed.
Despite recent progress in experimental measurements and theoretical
perturbative QCD analyses, our present knowledge of the partonic spin
structure of the nucleon is still incomplete. According to the conservation
law of the nucleon angular momentum, the nucleon spin is distributed among
quarks and gluons. The spin of the partonic constituents as well as their
orbital angular momentum contribute to the total nucleon spin of 1/2.
Recent experiments on polarised deep inelastic lepton-nucleon
scattering imply that quarks and anti-quarks carry only a small part of the
proton's spin (about 30$\%$) - less than half the prediction of relativistic
quark model (about 75$\%$). This result has stimulated theoretical activity
to understand the proton spin structure. A possible explanation for the
discrepancy may lie in a large gluon contribution, specially from the
small-Bjorken $x$ region. The present RHIC data cannot distinguish between
different (positive, negative and sign-changing) forms of the gluon
distributions $\Delta G(x)$ and therefore one cannot definitely determine
the quark and gluon contribution to the nucleon spin. Hence a primary goal
of the spin program is to determine the gluon polarisation $\Delta G$.\\ 
A knowledge of the low-$x$ behaviour of the unpolarised or polarised
nucleon structure functions enables the estimation of their Mellin moments
and hence the sum rules. This is particularly important in view of the
insufficient small-$x$ experimental data. Theoretical perturbative QCD
analysis is usually based on the evolution equations satisfied by the parton
densities. Interactions between quarks and gluons violate the Bjorken
scaling \cite{b1} and the parton distribution functions change with $Q^2$
according to the well-known Dokshitzer-Gribov-Lipatov-Altarelli-Parisi
(DGLAP) equation \cite{b2, b3, b4, b5}. 
The DGLAP equations can be solved with use of either the Mellin transform
or the polynomial expansion in the $x$-space. This approach requires
a knowledge of the initial parton densities at low-$Q^2$ scale for the
wide range of $x$-values. Input parametrisations are fitted to the
available experimental data. Standard DGLAP approach operates on the parton
densities $q$. Hence their moments, which are e.g. the contributions to
the proton's spin can be obtained by the integration of $q$ over $x$.
In this paper we propose an alternative approach, in which the main role is
played by truncated moments of the quark and gluon distribution functions.
The evolution equation for the $n$th truncated at $x_0$ moment
$\int\limits_{x_0}^1 dx\, x^{n-1}\, q(x,Q^2)$ has the same form as that for
the parton density itself with the modified splitting function
$P_{ij}'(n,x)= x^n P_{ij}(x)$ \cite{b6}.
The truncated moments approach, originated by S. Forte, L. Magnea,
A. Piccione and G. Ridolfi \cite{b6a}-\cite{b6c}, refers directly to
the physical values - moments (rather than to the parton densities), what
enables one to use a wide range of deep-inelastic scattering data in terms of
smaller number of parameters. In this way, no assumptions on the shape of parton
distributions are needed. Using the evolution equations for the truncated
moments one can also avoid uncertainties from the unmeasurable very small
$x\rightarrow 0$ region. This approach allows for direct study of the
behaviour of the truncated moments and can also be applied for determination
of the parton densities. The latter is a topic of our paper. 
Determination of the quark and gluon distributions from their truncated
moments  can be done very simply via differentiation of the moment with
respect to the point of truncation $x_0$. For practical purposes one can
also reconstruct the parameters in PDF parametrisation using Marquardt's
procedure. We describe this problem in detail and successfully apply our
method to recent HERMES and COMPASS spin-dependent valence quark data.

The content of this paper is as follows. In the next section we recall the
evolution equations for the truncated moments. Next, we present
a generalisation of the obtained equations for the double
truncated moments $\int\limits_{x_{min}}^{x_{max}} dx\, x^{n-1}\, q(x,Q^2)$.
Section 3 contains details for determination of the parton densities from
their truncated moments. Our approach is presented with help of a general
example for different input parametrisations of PDFs.
In Section 4 we reconstruct the valence quark densities from recent HERMES
and COMPASS data for their truncated first moments. In this way we can
compare our predictions for the reconstructed parton densities with the PDFs
fit and test the accuracy of our method. A summary and conclusions are given
in Section 5.

\section{Evolution equations for truncated moments}
\label{sec2}

Standard PQCD approach is based on DGLAP evolution equations for parton
densities \cite{b2, b3, b4, b5}. The DGLAP equations can be solved with use of
either the Mellin transform or the polynomial expansion in the $x$-space.
The differentio-integral Volterra-like evolution equations change after
the Mellin transform into simple differential and diagonal ones in the moment
space and can be solved analytically. Then one can again obtain the $x$-space
solutions via the inverse Mellin transform. The only problem is knowledge of
the initial moments - integrals over the whole region $0\leq x \leq 1$.
The lowest limit $x\rightarrow 0$, which implies that the invariant energy
$W^2=Q^2(1/x-1)$ of the inelastic lepton-hadron scattering becomes infinite,
will never be attained in experiments. Therefore it is very useful to study
the parton distributions only in a limited range of the Bjorken variable and
hence their moments truncated at low-$x_0$. The idea of truncated moments
was introduced and developed by S. Forte and his colleagues in
\cite{b6a}-\cite{b6c}.

In \cite{b6} we have derived the evolution equations for the truncated Mellin
moments of the parton densities. We have found that the truncated moments
obey the DGLAP-like evolution equation
\begin{equation}\label{eq.2.1}
\frac{d\bar{q}_n(x_0,Q^2)}{d\ln Q^2}=
\frac{\alpha_s(Q^2)}{2\pi}\; (P'\otimes \bar{q}_n)(x_0,Q^2),
\end{equation}
where
\begin{equation}\label{eq.2.2}
t = \ln(\frac{Q^2}{\Lambda^2}),
\end{equation}
$q(x,Q^2)$ is the parton distribution function and $\bar{q}_n(x_0,Q^2)$
denotes its $n$th moment truncated at $x_0$:
\begin{equation}\label{eq.2.3}
\bar{q}_{n}(x_0,Q^2)=\int\limits_{x_0}^1 dx\, x^{n-1}\, q(x,Q^2).
\end{equation}
A role of the splitting function plays $P'(n,z)$:
\begin{equation}\label{eq.2.4}
P'(n,z)= z^n\, P(z)
\end{equation}
and $\otimes$ abbreviates a Mellin convolution over $x$
\begin{equation}\label{eq.2.5}
(P \otimes f)\:\:(x,Q^2) = \int\limits_{x}^{1}\frac{dy}{y}\;
P\left ( \frac{x}{y}\right )\; f(y,Q^2)
\end{equation}
It can be shown, that the double truncated moments
\begin{equation}\label{eq.2.6}
\bar{q}_{n}(x_{min},x_{max},Q^2)=
\int\limits_{x_{min}}^{x_{max}} dx\, x^{n-1}\, q(x,Q^2)
\end{equation}
also fulfill the DGLAP-type evolution, namely
\begin{eqnarray}\label{eq.2.7}
\frac{d\bar{q}_n(x_{min},x_{max},Q^2)}{d\ln Q^2}=&&\nonumber\\
\frac{\alpha_s(Q^2)}{2\pi}\;
\int\limits_{x_{min}}^{1}\frac{dz}{z}\; P'(n,z)\;
\bar{q}_n\left ( \frac{x_{min}}{z}, \frac{x_{max}}{z}, Q^2 \right ) .
\end{eqnarray}
We would like to emphasize that although the central role in the QCD studies
play PDFs, dealing with their truncated moments can be also useful.
This approach has the following major characteristics:\\
- refers directly to the physical values - moments (not to the parton
densities), what enables one to use a wide range of experimental data in
terms of smaller number of parameters. In this way, no assumptions on the
shape of parton distributions are needed;\\
- allows one to study directly the evolution of moments and the scaling
violation;\\ 
- one can avoid uncertainties from the unmeasurable very small
$x\rightarrow 0$ and high $x\rightarrow 1$ region;\\
- the suitable evolution equations are exact and diagonal (there is no mixing
between moments of different orders);\\
- can be used for different approximations: LO, NLO, NNLO etc. and in the
polarised as well as unpolarised case.

Concluding, evolution equations for the truncated Mellin moments of the
parton densities (\ref{eq.2.1}), (\ref{eq.2.7}) can be an additional useful
tool in the QCD analysis of the nucleon structure functions.
In the next section we examine one of the possible application, which is
a reconstruction of the parton distributions.

\section{The application of the evolution equation for truncated moments}
\label{sec3}

The evolution equation (\ref{eq.2.1}) enables one to study the behaviour of
the truncated moments (\ref{eq.2.3}) within different approximations (LO,
NLO, NNLO etc.) and can be solved with use of standard methods of solving
the DGLAP equations. In this way one can study the evolution of the
truncated moments without making any assumption on the small-$x$ behaviour
of the parton densities themselves. One needs to know only the truncated
moments of the parton distributions at the initial scale $Q_0^2$ (e.g. from the
experimental data), what constrains a number of the input parameters.
The solutions for truncated moments can be used in the determination of
the parton distribution functions via differentiation
\begin{equation}\label{eq.3.1}
q(x,Q^2) = -x^{1-n}\:\frac{\partial\bar{q}_{n}(x,Q^2)}{\partial x},
\end{equation}
which results from (\ref{eq.2.3}). 
In order to reconstruct the parton densities from their truncated moments,
we proceed the following steps:\\
1. Preparing available experimental data for moments $\bar{q}_{n}(x_0,Q_1^2)$
as a function of $x_{min}\leq x_0\leq 1$ at the same scale $Q_1^2$.\\
2. Interpolation of the given data points into points which are Chebyshev
nodes. This allows us to use the Chebyshev polynomials technique for solving
the evolution equations.\\
3. Evolution of the truncated moments from $Q_1^2$ to $Q_2^2$ according to
(\ref{eq.2.1}) for different $x_{min}\leq x_0\leq 1$.\\
4. Reconstruction of the parton density $q(x,Q_2^2)$ from its truncated moment
at the same scale $Q_2^2$ applying Marquardt procedure to fit free
parameters.\\
Since numerical integration is more stable than numerical differentiation,
we perform the final fitting (in step 4.) with respect to moments and not to
their derivatives.\\

Let us test the above procedure on the nonsinglet function parametrised in
a general form:
\begin{equation}\label{eq.3.2}
q(x,Q_0^2) = N(\alpha, \beta, \gamma)\: x^{\alpha} (1-x)^{\beta}(1+\gamma x).
\end{equation}
Now we create `experimental data': for input parametrisation (\ref{eq.3.2})
we calculate truncated moments (\ref{eq.2.3}) at $Q_0^2$ and evolve them to
$Q^2$. Obtained results $\bar{q}_{n}(x_0,Q^2)$ are our starting point in the
above described 4-steps procedure. Finally we can compare the reconstructed
parton densities with the assumed form (\ref{eq.3.2}).\\
In our test we construct simulated data sets, what may seem to be a `toy'
proceeding. This approach has however a major advantage: knowing the result
`in advance', we are able test the accuracy of the reconstruction of the
parton densities. The presented evolution equations for the truncated
moments will be really useful in such QCD analyses, where we know the
moments (e.g. from direct measurements), while the PDFs are poorly known.\\
 
The results are presented in Fig.1. We performed our test using three
different input parametrisations $q(x,Q_0^2)$:
one nonsingular $(\sim const)$ and two singular $\sim x^{-0.4}$ and
$\sim x^{-0.8}$ as $x\rightarrow 0$.
\begin{figure}[ht]
\begin{center}
\includegraphics[width=85mm]{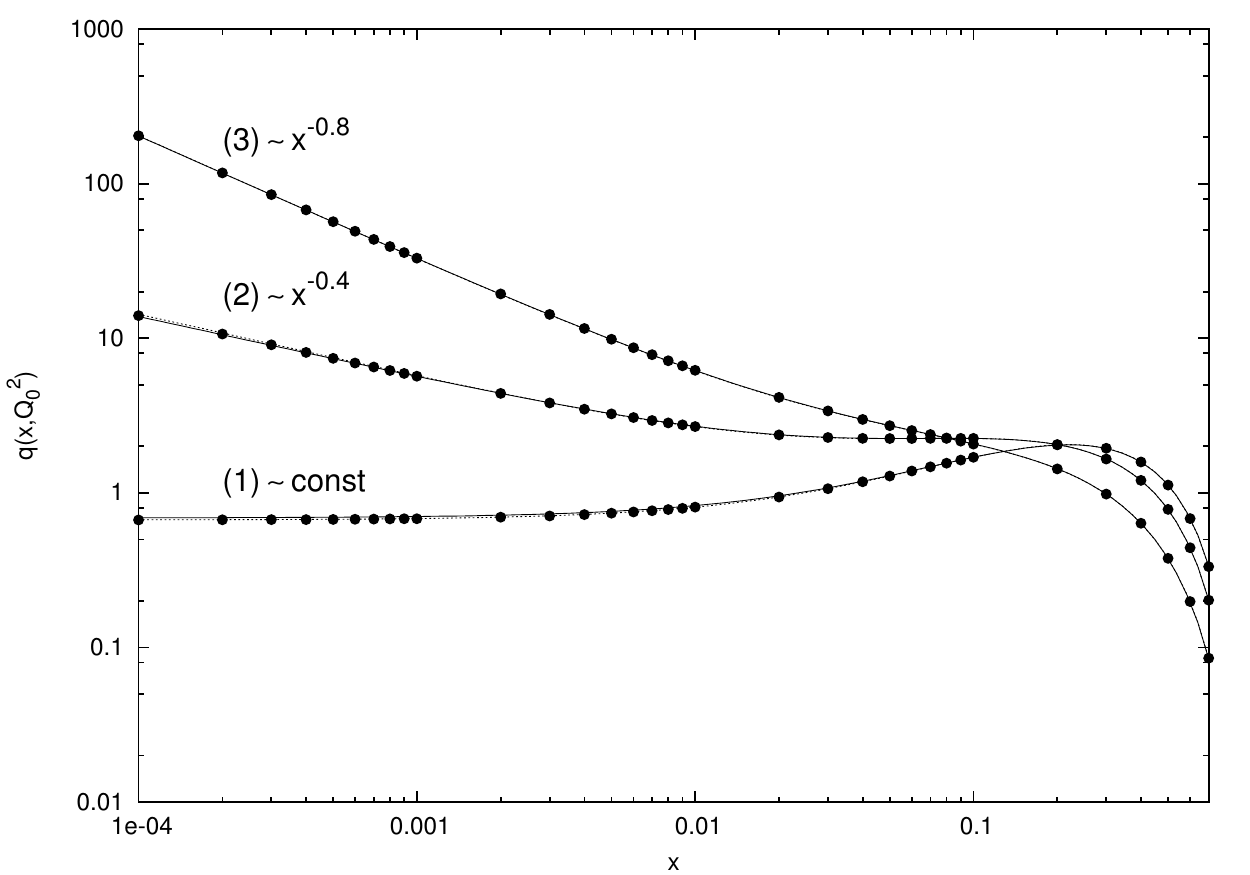}
\caption{Reconstruction of the initial parton density $q(x,Q_0^2)$
from its first truncated moment after back $Q^2$ evolution from
$Q^2=10\;{\rm GeV}^2$ to $Q_0^2=1\;{\rm GeV}^2$ (solid).
Comparison with the origin function $q(x,Q_0^2)$ (points). The dotted line
(mostly overlapped by the solid one) represents the reconstruction from the
limited $x$-region of available data $x_0\geq 0.01$.
Results are shown for 3 different small-$x$ behaviour of $q(x,Q_0^2)$:
(1)$(\sim const)$, (2)$\sim x^{-0.4}$ and (3)$\sim x^{-0.8}$.}
\end{center}
\end{figure}
One can see that our reconstruction is satisfactory independently on the
input parametrisation. The agreement is very good even for the limited region
of the data $x_0\geq 0.01$. It must be however emphasized that success of the
determination of the parton densities from their truncated moments depends
on the number of the fitted parameters and also on the available
data. A lack of data for the very small-$x$ is a constraint in the reliable
determination of $q(x,Q^2)$ in this region. More free parameters need more
data points and too many free parameters make a unique fit of the data not
possible. In our test we have fitted three parameters ($\alpha, \beta$ and
$\gamma$), obtaining almost the same values as those assumed in (\ref{eq.3.2}).
In the next section we shall reconstruct the valence quark distributions from
recent HERMES and COMPASS data. This needs to fit more parameters (minimum
three for u-quarks and three for d-quarks) and will be another test of the
accuracy of our method.

\section{Determination of the spin-dependent valence quark densities from
HERMES and COMPASS data for truncated first moments}
\label{sec4}

In the previous section we have presented the general idea of the
determination of the parton densities from their truncated moments.
Now, we will try to reconstruct the polarised valence quark densities from
recent HERMES \cite{b7} and COMPASS \cite{b8} data.
These data (moments) have not been obtained in direct measurements but with
use of a given PDF fit. Nevertheless, we can compare our predictions for
reconstructed parton densities with a given fit (input parametrisation) and
in this way test our method for a larger number of fitted parameters.
The results are shown in Figures 2-5.
\begin{figure}[ht]
\begin{center}
\includegraphics[width=85mm]{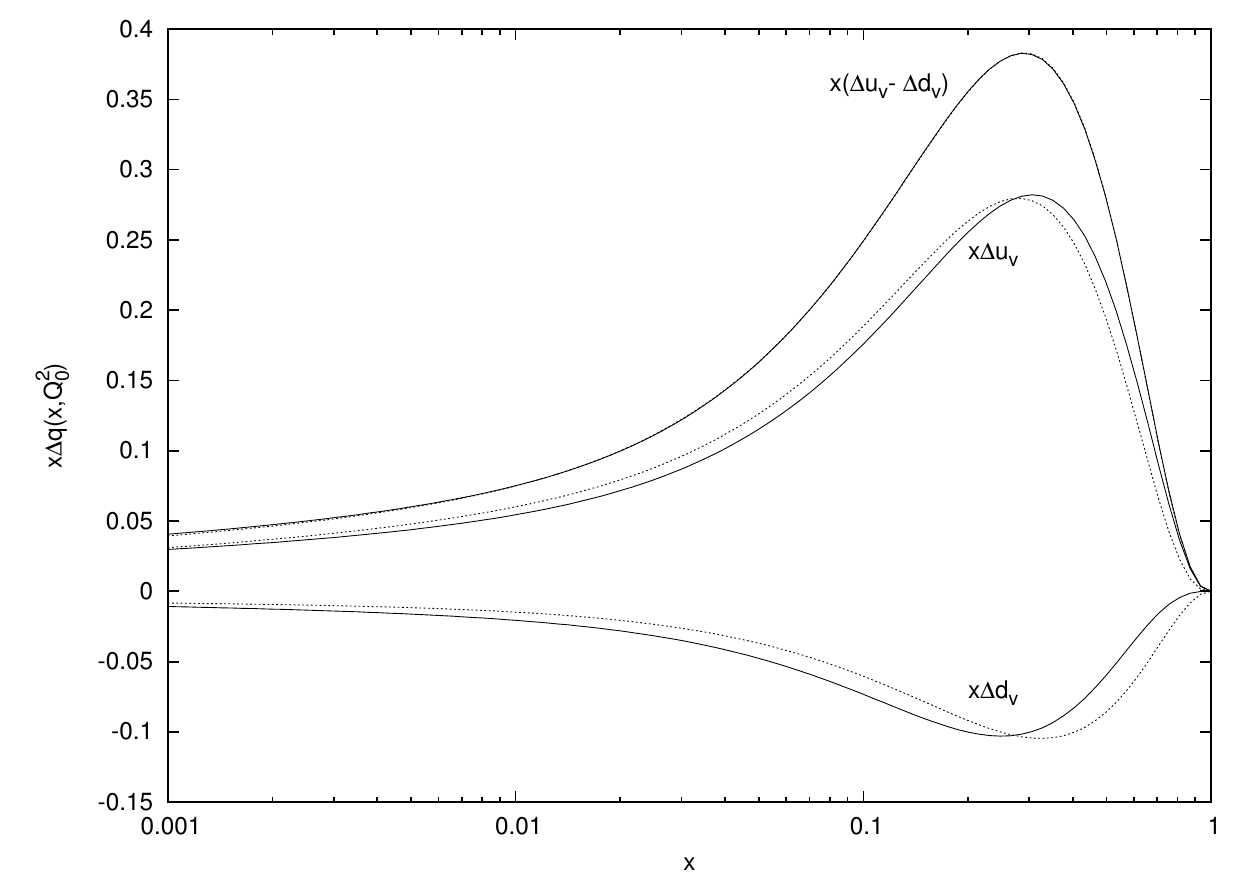}
\caption{Initial spin-dependent valence quark distributions
$x(\Delta u_{v}-\Delta d_{v})$, $x\Delta u_{v}$ and $x\Delta d_{v}$ at
$Q_0^2=4\;{\rm GeV}^2$:
dotted - reconstructed from HERMES data for the first truncated moment of
the nonsinglet polarised function $g_1^{NS}$ at $Q^2=5\;{\rm GeV}^2$ \cite{b7},
solid - original BB fit \cite{b9}. Plots for $x(\Delta u_{v}-\Delta d_{v})$
overlap each other.}
\end{center}
\end{figure}
\begin{figure}[ht]
\begin{center}
\includegraphics[width=85mm]{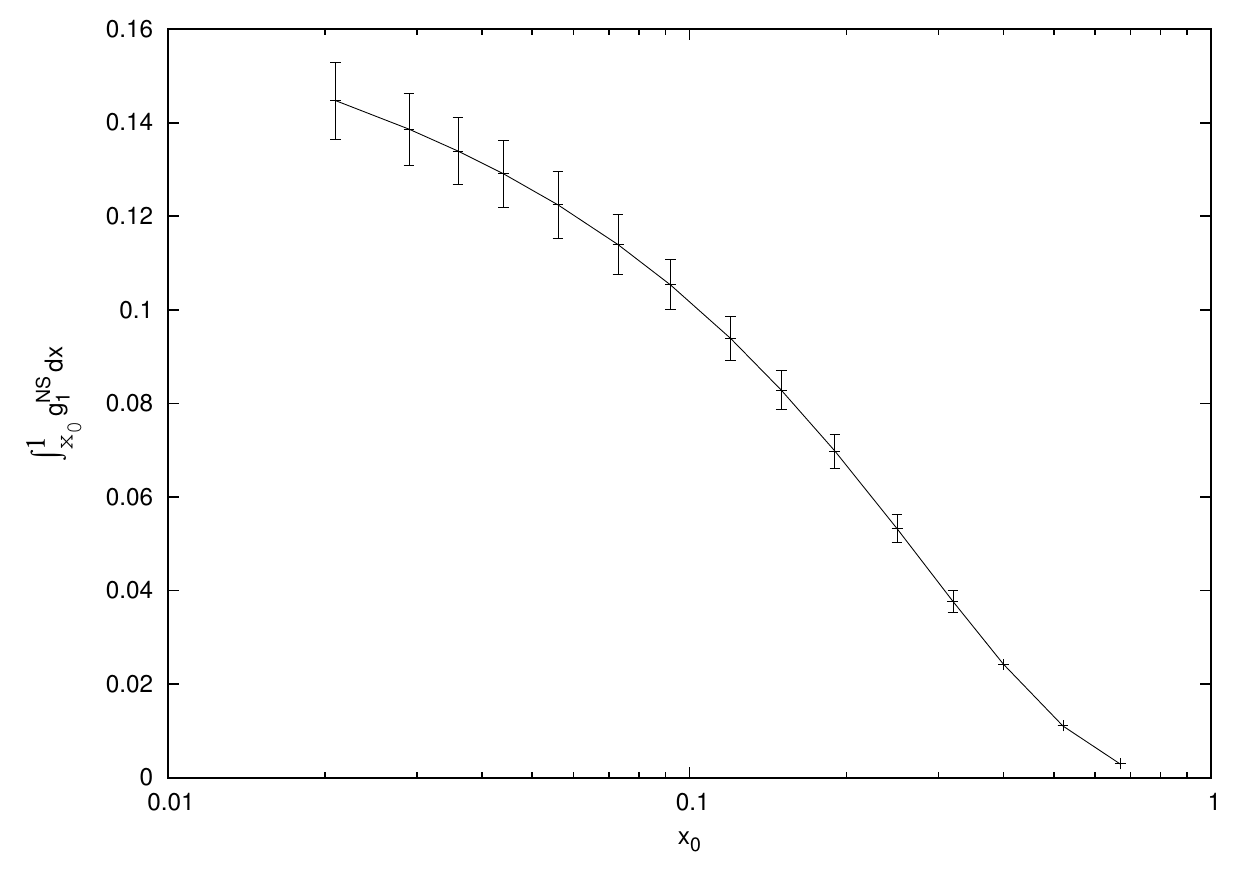}
\caption{The first truncated moment of the nonsinglet polarised structure
function $g_1^{NS}$ vs the truncation point $x_0$, calculated from the
reconstructed fit (solid). $Q^2=5\;{\rm GeV}^2$.
Comparison with HERMES data \cite{b7} based on BB fit \cite{b9} (points with
error bars).}
\end{center}
\end{figure}
\begin{figure}[ht]
\begin{center}
\includegraphics[width=85mm]{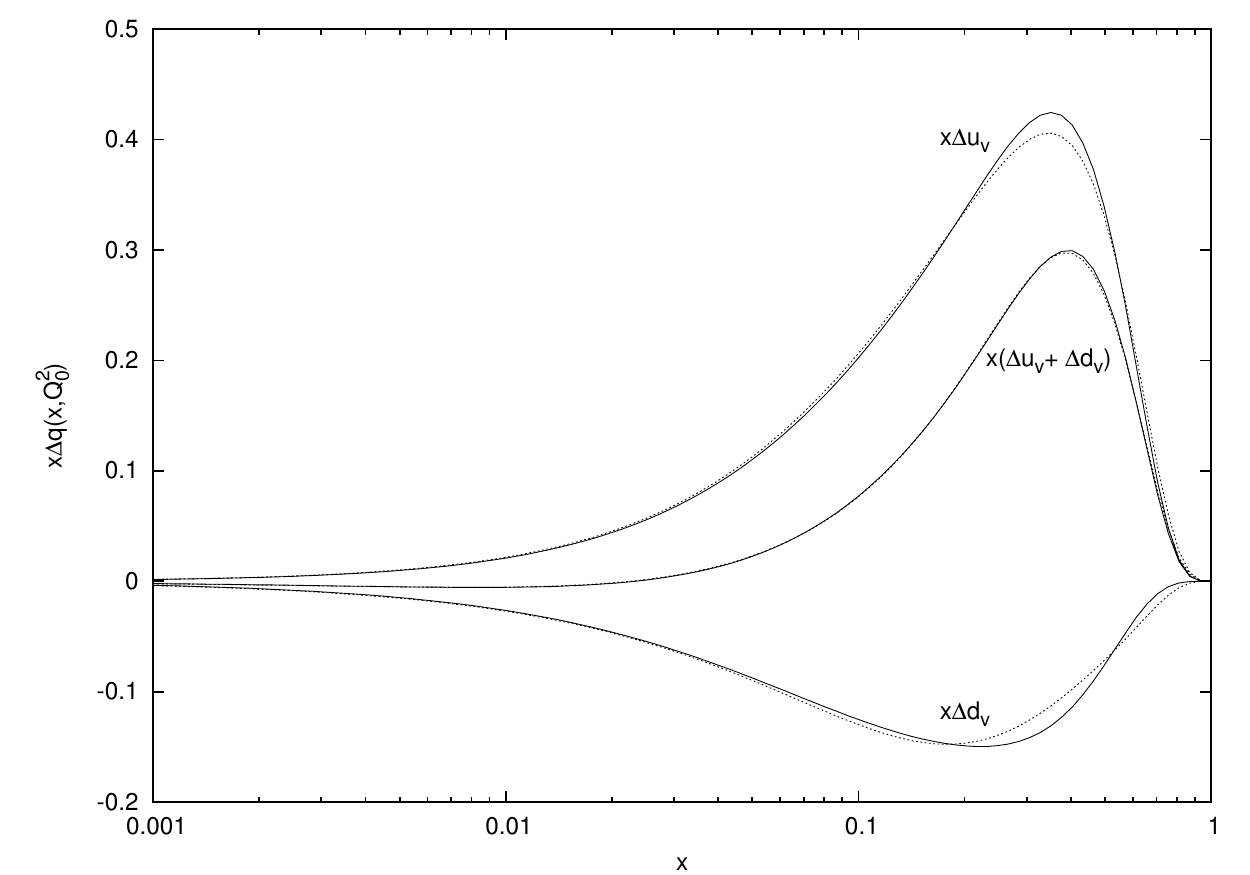}
\caption{Initial spin-dependent valence quark distributions
$x(\Delta u_{v}+\Delta d_{v})$, $x\Delta u_{v}$ and $x\Delta d_{v}$ at
$Q_0^2=0.5\;{\rm GeV}^2$:
dotted - reconstructed from COMPASS data for the first truncated moment of
the function $\Delta u_{v}+\Delta d_{v}$ at $Q^2=10\;{\rm GeV}^2$
\cite{b8}, solid - original DNS fit \cite{b10}.
Plots for $x(\Delta u_{v}+\Delta d_{v})$ overlap each other.}
\end{center}
\end{figure}
\begin{figure}[ht]
\begin{center}
\includegraphics[width=85mm]{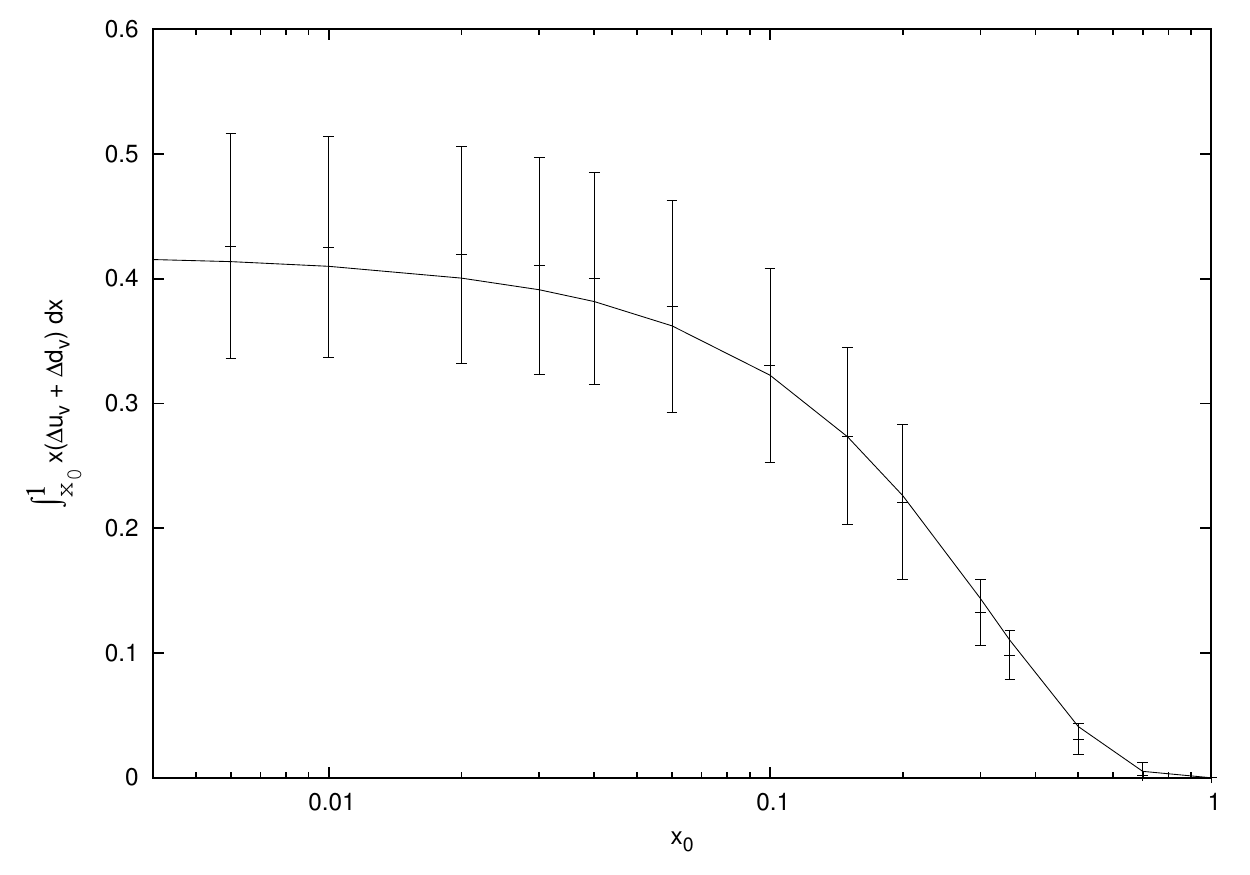}
\caption{The first truncated moment of the function $\Delta u_{v}+\Delta d_{v}$
vs the truncation point $x_0$, calculated from the reconstructed fit (solid).
$Q^2=10\;{\rm GeV}^2$. Comparison with COMPASS data \cite{b8}
based on DNS fit \cite{b10} (points with error bars).}
\end{center}
\end{figure}

In Fig. 2 we plot the initial spin-dependent valence quark distributions
$x\Delta u_{v}(x,Q_0^2)$ and $x\Delta d_{v}(x,Q_0^2)$, reconstructed from
the HERMES data \cite{b7} for the first truncated moment of the nonsinglet
polarised function
\begin{equation}\label{eq.4.1}
g_1^{NS} = \frac{1}{6} (\Delta u_{v}-\Delta d_{v}).
\end{equation}
We compare our results to the original Bl\"umlein - B\"ottcher (BB) fit
\cite{b9}.
Fig. 3 shows $x_0$ dependence of the first truncated moment
\begin{equation}\label{eq.4.2}
\bar{g}_1^{NS}(x_0, Q^2) = \int_{x_0}^{1} g_1^{NS}(x,Q^2) dx
\end{equation} 
calculated from the reconstructed fit, compared with the experimental data.
Figs. 4 and 5 contain the same analysis as Figs. 2 and 3 respectively, but
for COMPASS $\Delta u_{v}+\Delta d_{v}$ data \cite{b8} with the
original de Florian, Navarro, Sassot (DNS) fit \cite{b10}.
  
One can see a satisfactory agreement between the reconstructed fits and
experimental data. Reconstructed combined functions
$x(\Delta u_{v}-\Delta d_{v})$ and $x(\Delta u_{v}+\Delta d_{v})$ overlap
HERMES and COMPASS results, respectively. For the extracted valence quark
densities alone the agreement is worse but still acceptable. We have found,
however, that these fits are not unique and equally good agreement with the
data can be obtained with use of the other (not only BB and DNS respectively)
sets of free parameters. When the number of adjustable parameters is large
($>3$) and there are no experimental points from the low-$x$
region $x<0.001$, one cannot distinguish which fit is the best one. Only an
additional constraint for small-$x$ behaviour of the parton densities makes
the fit procedure more reliable. In our test we found the best HERMES fit
taking into account also the small-$x$ $(x<0.01)$ $g_1^{NS}$ data.
Concluding, even for the large number of adjustable parameters (6 for HERMES
and 8 for COMPASS data), the presented method of reconstruction can be
a hopeful tool for determining parton densities from experimental results for
their truncated moments. 

\section{Summary}
\label{sum}

We have shown how to determine the parton densities from their first truncated
at $x_0$ Mellin moments. The method is based on the recently derived evolution
equations for $n$th truncated moments $\bar{q}_n(x_0,Q^2)$. After evolution of
the $\bar{q}_1$ results from a given scale $Q^2$ to the initial one $Q_0^2$,
one can reconstruct the input quark and gluon distributions. In our analyses
we have used different simulated data sets in order to test the accuracy of
the reconstruction.
In our first test we have fitted three parameters from the general form of
the nonsinglet input function, obtaining a very good agreement with assumed
original parametrisations. We have found, that for small number of the fitted
parameters ($\leq 3$), the reconstruction is satisfactory independently on the
small-$x$ behaviour of the assumed input and  even for the limited region of
the data $x_0\geq 0.01$. Next we have applied
this technique to the experimental HERMES and COMPASS data for the polarised
valence quarks.
In this way we have tried to reconstruct the original fits with
larger number of free parameters (6 and 8 respectively). The results of the
reconstruction has been again satisfactory. We have found however, that for
larger number of adjustable parameters the obtained fits are not unique.
We would like to emphasize that success of the determination of the parton
densities from their truncated moments depends on the number of the fitted
parameters and also on $x$-region of the available experimental data. An
additional knowledge of the small-$x$ behaviour of the parton densities,
based either on the experimental data or the theoretical expectations,
can make the fit procedure more reliable. Indeed, we have found the
best HERMES fit taking into account also the small-$x$ $g_1^{NS}$ data.

Concluding, the presented evolution equations for the truncated moments
allowing for a direct study of the $Q^2$ - dependence of the moments, offer
a new, promising tool towards improving our knowledge of the unpolarised
and polarised partonic structure of the nucleon.


\begin{thebibliography}{99}

\bibitem{b1}
J.D. Bjorken, Phys. Rev. \textbf{179}, 1547 (1969)
\bibitem{b2}
V.N. Gribov, L.N. Lipatov, Sov. J. Nucl. Phys. \textbf{15}, 438 (1972)
\bibitem{b3}
V.N. Gribov, L.N. Lipatov, Sov. J. Nucl. Phys. \textbf{15}, 675 (1972)
\bibitem{b4}
Yu.L. Dokshitzer, Sov. Phys. JETP \textbf{46}, 641 (1977)
\bibitem{b5}
G. Altarelli, G. Parisi, Nucl. Phys. B \textbf{126}, 298 (1977)
\bibitem{b6}
D. Kotlorz, A. Kotlorz, Phys. Lett. B \textbf{644}, 284-287 (2007),
\href{http://www.arxiv.org/abs/hep-ph/0610282}{[hep-ph/0610282]}
\bibitem{b6a}
S. Forte, L. Magnea, Phys. Lett. B \textbf{448}, 295 (1999)
\href{http://www.arxiv.org/abs/hep-ph/9812479}{[hep-ph/9812479]}
\bibitem{b6b}
S. Forte, L. Magnea, A. Piccione, G. Ridolfi, Nucl. Phys. B \textbf{594}, 46
(2001)
\href{http://www.arxiv.org/abs/hep-ph/0006273}{[hep-ph/0006273]}
\bibitem{b6c}
A.Piccione, Phys. Lett. B \textbf{518}, 207 (2001)
\href{http://www.arxiv.org/abs/hep-ph/0107108}{[hep-ph/0107108]}
\bibitem{cheb}
S.E. El-gendi, Comput. J. \textbf{12}, 282 (1969)
\bibitem{b7}
HERMES Collaboration, A. Airapetian et al., Phys. Rev. D \textbf{75}, 012007
(2007),
\href{http://www.arxiv.org/abs/hep-ex/0609039}{[hep-ex/0609039]}
\bibitem{b8}
COMPASS Collaboration, M. Alekseev et al., Phys. Lett. B \textbf{660},
458-465 (2008),
\href{http://www.arxiv.org/abs/arXiv:0707.4077}{[arXiv:0707.4077]}
\bibitem{b9}
J. Bl\"umlein, H. B\"ottcher, Nucl. Phys. B \textbf{636}, 225-263 (2002),
\href{http://www.arxiv.org/abs/hep-ph/0203155}{[hep-ph/0203155]}
\bibitem{b10}
D. de Florian, G.A. Navarro, R. Sassot, Phys. Rev. D \textbf{71}, 094018
(2005),
\href{http://www.arxiv.org/abs/hep-ph/0504155}{[hep-ph/0504155]}

\end{thebibliography}
\end{document}